\documentclass{article}
\usepackage{amsfonts}
\usepackage{amssymb}
\usepackage{amsmath}
\usepackage{amsthm}
\usepackage{undertilde}
\usepackage[english]{babel}
\begin{document}

\title{\bf Revisiting the relaxation of constraints\\ in gauge theories}

\author{Alexey Golovnev${}^{1}$, Kirill Russkov${}^{2,3}$\\
{\small ${}^{1}${\it Centre for Theoretical Physics, the British University in Egypt,}}\\ 
{\small {\it BUE 11837, El Sherouk City, Cairo Governorate, Egypt}}\\
{\small agolovnev@yandex.ru}\\
{\small ${}^{2}${\it Department of High Energy and Elementary Particle Physics, Saint Petersburg State University,}}\\ 
{\small {\it Ulianovskaya, 1, Stary Petergof,  Saint Petersburg, Russia}}\\
{\small ${}^{3}${\it Petersburg Nuclear Physics Institute of National Research Centre “Kurchatov Institute”,}}\\ 
{\small {\it Gatchina 188300, Russia}}\\
{\small kerill.russ42alex@gmail.com}
}
\date{}

\maketitle

\begin{abstract}

Recently, there were works claiming that path integral quantisation of gauge theories necessarily requires relaxation of Lagrangian constraints. As has also been noted in the literature, it is of course wrong since there perfectly exist gauge field quantisations respecting the constraints, and at the same time the very idea of changing the classical theory in this way has many times appeared in other works. On the other hand, what was done in the path integral approach is fixing a gauge in terms of zero-momentum variables. We would like to show that this relaxation is what normally happens when one fixes such a gauge at the level of action principle in a Lagrangian theory. Moreover, there is an interesting analogy to be drawn. Namely, one of the ways to quantise a gauge theory is to build an extended Hamiltonian and then add new conditions by hand such as to make it a second class system. The constraints' relaxation occurs when one does the same at the level of the total Hamiltonian, i.e. a second class system with the primary constraints only.

\end{abstract}

\section{Introduction}

In recent preprints \cite{strin, strgr, strelm, strcons}, an idea of relaxing the Lagrangian constraints in gauge theories has been put forward. The timeflow of these claims was as follows. The initial paper \cite{strin} dealt with the problem of time in quantum cosmology, or more precisely with the rather natural fact that its Hamiltonian vanishes on-shell. For sure, relaxing the first Friedmann equation does allow us to make the on-shell Hamiltonian non-zero. Then the next papers \cite{strgr, strelm} discussed relaxing the constraints in full General Relativity and in electromagnetism respectively, while the last paper in the series \cite{strcons} continued considering some cosmological implications.

Of course, we should try all possible theoretical ideas, and the constraints' relaxation is no exception to that. However, these papers claimed it to be a necessity \cite{strin, strgr} of quantum physics based on a na{\" i}ve approach to path integrals as if, modulo summing over the gauge copies, those were always well-defined and mathematically clean which would probably entail a perfect quantum gravity theory with flawless background independence and everything, let alone a consistent  mathematical description of all our quantum field theories, nonperturbatively. This is not the case.

Another thing to mention is that, with the poor level of mathematical rigour available in quantum field theory, we do know how to quantise gauge theories without relaxing the constraints, and even in a few different approaches. We recommend a very nice paper \cite{Kamen} and references therein for a review on the topic. Note that another point they make \cite{Kamen} is also worth paying attention to. Namely, these ideas of relaxing the constraints, though admitting that it is a modification at the level of the classical theory, have been appearing many times in the literature and date back to old works by V.A. Fock and E.C.G. St{\" u}ckelberg; see the review and references in the paper \cite{Kamen}.

We find it interesting to add yet another aspect to this theme. The paper \cite{Kamen} is written in the Hamiltonian language or, one can say, in terms of the first-order action functionals. We show that, starting from the proper Lagrangian description of gauge theories, the relaxation happens\footnote{As we will see below, it is only for those of them in which the "gauge hits twice" \cite{Teitelboim}.} when we take the primary constraints only and fix the gauge by adding conditions which are of second class with them. 

\section{The simple case of electrodynamics in vacuum}

Let us first discuss the most classical field theory, that of electrodynamics. Its quantisation is very well-known, though the details can always be discussed \cite{DrGl}. For the sake of simplicity, we take it in vacuum. The dynamical variable is a vector field $A^{\mu}$ and the field strength is defined as
\begin{equation}
\label{strength}
F_{\mu\nu} = \partial_{\mu} A_{\nu} -  \partial_{\nu} A_{\mu}
\end{equation}
satisfying the Maxwell equation\footnote{To be precise, this is one half of the Maxwell equations, those which would be sourced by electric charges and their currents, had we not taken the model in vacuum. The representation (\ref{strength}) by itself is the solution for another half.} $\partial_{\nu} F^{\nu\mu}=0$.

{\bf The standard description}. The Lagrangian density and the total Hamiltonian density are
\begin{equation}
\label{elmst}
{\mathcal L}=-\frac14 F_{\mu\nu}^2=\frac12 F^2_{0i} - \frac14 F_{ij}^2, \qquad {\mathcal H}_T = \frac12 \pi^2_i + \pi_i \partial_i A_0 +  \frac14 F_{ij}^2 + \lambda \pi_0
\end{equation}
where the Lagrange multiplier $\lambda$ imposes the primary constraint of $\pi_0=0$. Taking the usual definition of spatial momenta, $\pi_i=\frac{\partial \mathcal L}{\partial {\dot A}_i}$, we immediately see that finally both pictures produce the same equations
$$\partial_i F_{i0}=0,\qquad {\dot F}_{0i} - \partial_j F_{ji} =0, \qquad \pi_0=0, \qquad \pi_i=F_{0i}$$
the first of which is the Gau{\ss} law. The arbitrary Lagrange multiplier then represents the gauge freedom of choosing $A_{\mu}$ for a given field strength $F_{\mu\nu}$.

We prefer to treat the primary constraint as yet another equation, on par with the dynamical equations of motion \cite{meHam}. In the common way of presenting the constrained Hamiltonian dynamics, the constraint would be given as a "weak equality" $\pi_0\approx 0$ and taken as just an initial condition, and the Gau{\ss} law would be derived by its preservation in time as $\partial_i \pi_i \approx 0$, i.e. the secondary constraint. Both viewpoints are possible, and one would anyway need to find all the constraints when, due to any reason, the knowledge of a full set of possible initial conditions is required. 

Note though that the initial idea of Dirac was different. In the first paper \cite{inDirac} on this topic, instead of $=$ and $\approx$ signs, he used $\equiv$ and $=$ respectively. He considered arbitrary independent variations of $q$, $\dot q$, $p$ variables (coordinate, velocity and momentum) and named an equation weak if it was valid to the zeroth order only, i.e. if it was immediately violated by a generic variation, and strong if it was correct up to the first order, at least, or maybe even exact. For example, the Lagrangian function $L\equiv L(q,\dot q)$ was assumed to be given as a strong equality, while the definition of momenta $p=\frac{\partial L}{\partial \dot q}$ as a weak one. Then, eight years later, in a paper\footnote{It was published together with his Hamiltonian form of general relativity, see below.} \cite{Dirac} of the same title but in another journal, he changed the notation to the familiar $=$ and $\approx$ one and claimed to have put "the method in a more direct and practical form". The new motivation for defining a weak equality was that it is just a reminder of the fact that any Poisson bracket must be calculated before using equations which restrict the values of coordinates and momenta. In reality, there is nothing new in that. For example, a derivative of a function must also be taken before evaluating it at a particular point.

{\bf Extended Hamiltonian and gauge fixing}. A common opinion \cite{Teitelboim}, dating back to Dirac himself, is that one should explicitly add all the constraints to the Hamiltonian (\ref{elmst}),
\begin{equation}
\label{elmext}
{\mathcal H}_E  = \frac12 \pi^2_i + \pi_i \partial_i A_0 +  \frac14 F_{ij}^2 + \lambda \pi_0 +\tilde{\lambda} \partial_i \pi_i
\end{equation}
and work with this so-called extended Hamiltonian (\ref{elmext}). The dynamical, i.e. spatial divergenceless, modes remain unmodified. However, the rest of it gets curious modifications \cite{meHam, mesing}. One of the Hamiltonian equations is changed: 
$${\dot A}_i = \pi_i + \partial_i A_0 - \partial_i \tilde\lambda,\qquad  {\dot A}_0=\lambda, \qquad {\dot\pi}_i=\partial_j F_{ji},\qquad \pi_0=\partial_i \pi_i=0$$ 
while both Lagrange multipliers are arbitrary. We see that the definition of momentum is lost in its spatial longitudinal part, and the quantity $\partial_i (F_{0i} - \pi_i)$ becomes pure gauge. The usual equations outside the transverse sector, that is the Gau{\ss} law, are not there unless we assume that the electric field is now given by $\pi_i$ while $F_{0i}$ has partly lost its physical meaning.

What they do when proving the "Dirac's conjucture" \cite{conjDirac} is artificially enhancing the set of gauge symmetries and losing some of the standard definitions of momenta which remain valid only in a special gauge given by a choice equivalent to $ \tilde\lambda=0$. With the new definition of gauge invariance, the gauge invariant information is indeed unchanged by extending the Hamiltonian. From the Lagrangian point of view, only the case of $\tilde\lambda=0$ is unmodified.

Actually, one of the ways to quantise a gauge model is to take the extended Hamiltonian and fix a gauge \cite{Teitelboim}, i.e. to add new constraints which would be a set of second class with the real ones. One can even do so without violating the Lagrangian equations. For that we need to choose a gauge with $\tilde\lambda=0$. One option would be to take $A_0=\partial_i A_i=0$ (recall that we are in vacuum):
$${\mathcal H}_{\mathrm{fixed}}  = \frac12 \pi^2_i + \pi_i \partial_i A_0 +  \frac14 F_{ij}^2 + \lambda \pi_0 +\tilde{\lambda} \partial_i \pi_i + \utilde{\lambda} A_0 + \utilde{\tilde\lambda} \partial_i A_i .$$
One can check that, given suitable boundary conditions at spatial infinity, there is no gauge freedom left, for the equations imply $\lambda=\utilde{\lambda}=\bigtriangleup\tilde\lambda=\bigtriangleup\utilde{\tilde\lambda}=0$, and the usual Maxwell equations for $F_{\mu\nu}$ are there.

{\bf Relaxation of constraints}. Finally, we come to the point of relaxing Lagrangian constraints \cite{strelm}. One adds a constraint like $A_0=0$, or with an arbitrary spacetime function $A_0=f(x)$, to the total Hamiltonian:
\begin{equation}
\label{relelm}
{\mathcal H}_{\mathrm{relaxed}}  = \frac12 \pi^2_i + \pi_i \partial_i A_0 +  \frac14 F_{ij}^2 + \lambda \pi_0 + \utilde{\lambda} A_0 ,
\end{equation}
with the last constraint possibly changed to $A_0-f(x)$ instead of simple $A_0$. Since the Poisson bracket of $\pi_0$ and $A_0$ is nowhere zero, their preservation in time fixes the value of the Lagrange multipliers and no new constraints are generated.

One can do the same by simply substituting $A_0=f(x)$ into the Lagrangian. In particular, the $A_0=0$ option is very easy to deal with because $F_{0i}={\dot A}_i$ then. In any case, we get ${\dot F}_{0i} - \partial_j F_{ji} =0$ and no Gau{\ss} law constraint. Therefore, we can write the equations as
\begin{equation}
\label{releq}
\partial_{\nu} F^{\nu\mu}=\rho \delta^{\mu}_0
\end{equation}
with an arbitrary function $\rho(x)$. Taking a divergence, we get the only restriction on it as $\dot\rho=0$. This is simply the consequence of the necessary charge conservation. Since there are no currents associated with this modification, the charge density cannot depend on time. 

In other words, the classical theory has been changed by a possible addition of a fixed static background of electric charges for every universe described by this theory. There are two points to make. First, there is no need to consider the whole story anew in this way, for one can just take the usual electromagnetism with an external source. Otherwise, it is a Lorentz-violating idea, to go for treating it as something fundamental instead of being an external background put by hand. Indeed, the component $A_0$, or presence of charge density in absence of currents, depends on the choice of an inertial frame. 

{\bf A remark}. If we had fixed the gauge of $A_0=0$ in the field equations, the result would have been
$${\ddot A}_i =\partial_j F_{ji}, \qquad \partial_i {\dot A}_i =0.$$
Due to the last equation, the longitudinal component could then be counted as one half dynamical degree of freedom. Actually, this is nothing but the remaining gauge freedom of $\delta A_{\mu} =\partial_{\mu} \phi(x_i) $. However, when imposed directly inside the action functional, it leads to ${\ddot A}_i =\partial_j F_{ji}$ being the only equation. In this case, we can only conclude that $\partial_i {\ddot A}_i =0$. In other words, all three modes are fully dynamical, i.e. one full new degree of freedom, only one half of which belongs to what was the gauge freedom. Formally, for the gauge invariant variables $F_{\mu\nu}$, we have got one half new dynamical degree of freedom (\ref{releq}) since the function $\rho(x_i)$ acts as one new Cauchy datum.

\section{On other gauge choices}

What we have seen above is that the relaxation procedure depends on the choice of a reference frame. Adding an effective electric charge distribution without adding any electric currents does depend on the frame. It is no surprise because the Hamiltonian approach itself is formulated in a preselected time variable. Another feature of relaxation is that it depends on the choice of variables. 

Suppose, we have changed variables in the Maxwell Lagrangian to ${\mathfrak A}_0 = A_0 + f(A_i)$ and $A_i$ themselves unaffected. Then the relaxation gauge condition of ${\mathfrak A}_0=0$ corresponds to $A_0=-f(A_i)$ in terms of initial variables. Even though such gauges were dismissed by the authors of relaxation \cite{strin}, it is a perfectly legitimate gauge choice, too. Below we will first consider these options, i.e. the second-class ones with the primary constraint, then briefly comment on other possible gauge choices.

{\bf Keeping up with relaxation of Lagrangian constraints}. Let us assume that, on top of the primary constraint $\pi_0=0$, we have also added our artificial constraint $A_0 + f(A_i)=0$ to the total Hamiltonian of electrodynamics (\ref{elmst}). Since those two constraints are second class, we still get the same result of no secondary constraint appearing. In other words, the Gau{\ss} law has been cancelled. However, one can easily see that, for a nontrivial function $f$ in the constraint, the Hamiltonian equations become a bit more complicated, also in their final shape for the field strength (\ref{strength}).

It makes sense. Indeed, we have got a system of Lagrangian equations which was of at most second order, but also with a lower derivative part in it. Relaxing all the lower-order information, like the Gau{\ss} law, is a well-defined statement. However, what remains after that depends on the way the system has been written, for one could take a linear combination of equations. One might demand removing all possible linear combinations of a lower derivative order while keeping its linear complement to the full system. The latter is not uniquely defined.

Indeed, as far as we have substituted a gauge of $A_0=-f(A_i)$ into the Lagrangian of the Maxwell action functional $S$, we derive the equation of motion
$$\frac{\delta S}{\delta A_i} - \frac{\delta S}{\delta A_0}\cdot \frac{\partial f}{\partial A_i} =0$$
with no independent condition on $\frac{\delta S}{\delta A_0}$. Therefore, we still get a possible distribution of effective electric charges $\rho$, though this time it comes with electric currents $j^i$ given by $\rho \frac{\partial f}{\partial A_i} $, and with the restriction of charge conservation as always.

{\bf Other options}. As should be obvious by now, the effect of relaxing the constraints is an artifact of fixing a gauge directly in the action principle, or in the total Hamiltonian, when the gauge condition is of second class with the primary constraint. It simply prevents the secondary constraint from being generated. Another option is to choose a gauge without this side effect. Since space is no different from time, putting a condition of, say, $A_1=0$ into the Lagrangian leaves us with remnant gauge freedom of $\delta A_{\mu}=\partial_{\mu} \phi (t,x_2,x_3)$, with no dependence on $x_1$. In perturbation theory, this unwanted freedom can be taken care of by boundary conditions at spatial infinity of $x_1$.

Since we usually need to respect spatial isotropy, it would be nice to consider the Coulomb gauge condition of $\partial_i A_i=0$ in more detail. Unlike the temporal one, the spatial longitudinal condition is of first class with respect to the primary constraint, and therefore the secondary constraints appear, $\partial_i \pi_i =0$ for preservation of the natural primary constraint and $\bigtriangleup A_0=0$ for preservation of the hand-made one.

Using integrations by parts, the Coulomb gauge condition allows one to bring the Lagrangian density to the form of ${\mathcal L}=\frac12 \left({\dot A}_i^2 - (\partial_i A_j)^2 + (\partial_i A_0)^2\right)$ with equations
$$\bigtriangleup A_0 =0, \qquad \square A_i =0,$$
while using this gauge in equations only produces
$$\bigtriangleup A_0 =0, \qquad \square A_i =\partial_i {\dot A}_0.$$
The difference disappears in perturbation theory when we solve $\bigtriangleup A_0 =0$ simply as $A_0=0$, and in any case it is inside the remnant symmetry of  $\delta A_{\mu}=\partial_{\mu} \phi$ with $\bigtriangleup\phi=0$.

The gauges which do not immediately fix $A_0$ are better from the practical point of view. Normally, they can be safely imposed directly inside the action, at least for the aims of perturbation theory. This fact is often forgotten but, in general, doing it with arbitrary gauges does change the theory at hand. One can, for example, discuss it in the context of cosmological perturbation theory \cite{Lagos}. On the other hand, one can easily find a gauge which behaves even worse than the temporal one. The typical example, much used in quantum field theory, is the Lorenz gauge, $\partial_{\mu}A^{\mu}=0$. The remaining gauge freedom is a full-fledged fake degree of freedom, $\square\phi=0$, while its substitution into the Maxwell action makes a disaster of ${\mathcal L}=\frac12 (\partial_{\mu} A_{\nu})^2$ upon integrations by parts.

\section{The general arguments for mechanical systems}

An important point about electrodynamics above is that its gauge freedom "hits twice" \cite{Teitelboim}. This expression means that one gets twice as many first-class constraints as there are independent gauge symmetries, which in turn leads to elimination of two dynamical modes for every such symmetry. Therefore, with only one gauge symmetry of electrodynamics, there are $4-2=2$ dynamical modes. The gauge transformation mixes different types of derivatives. The $A_0$ component lacks the time derivative in the action, while another component, $A_i$, lacks it own $\partial_i$ derivative there. Therefore, the primary constraint is not a gauge generator in itself, unless we go for the extended Hamiltonian, and it generates the secondary constraint which can finally be translated into the Lagrangian one. Only their combination then serves as a generator of gauge transformations in the initial meaning \cite{Castellani}.

A trivial exception to this situation is when a Lagrangian does not depend on one of its formal arguments at all. Then this particular variable is fully free, and vanishing of its momentum does not produce any new constraint. Making the extra variable be equal to any fixed function of spacetime is an absolutely full gauge fixing. For example, rewriting general relativity in terms of an orthonormal tetrad produces local Lorentz symmetry which does not hit twice. One can choose a particular tetrad and act with a Lorentz matrix on it. If everything was written in terms of the metric, this matrix totally drops out from the action. At the same time, if this drop-out is up to a total derivative term only, like in teleparallel gravity, the formal Hamiltonian procedure might go in a complicated way \cite{Lor}.

Let us now argue that the situation we have found in the case of electrodynamics is very general for any gauge symmetries hitting twice\footnote{Somewhat artificially, one can construct systems in which it hits more than twice, for example a Lagrangian \cite{Castellani} of $L={\dot x}{\dot z} + yz$ where the primary constraint $p_y=0$ entails two secondary (or, one can say, a secondary and a tertiary) constraints $z=0$ and $p_x=0$, all of first class. Fixing $y=0$ inside the action removes them both. Up to a boundary term, it was a gauge symmetry of $\delta x= \epsilon (t)$, $\delta y=\ddot\epsilon (t)$.}. For simplicity, consider a mechanical model 
\begin{equation}
\label{meclag}
L=L(q, {\dot q}, s)
\end{equation}
with two variables $q(t)$ and $s(t)$, though it wouldn't be difficult to consider them multidimensional. Assume also that $\frac{\partial L}{\partial \dot q}\neq 0$ and there is a gauge freedom under transformations of the form
$$\delta s(t) = f(q,s) \cdot {\dot \epsilon} (t) + h(q,s) \cdot  \epsilon (t), \qquad \delta q(t) = g(q,s) \cdot \epsilon (t)$$
with fixed functions $f,h,g$ nontrivially mixing $\delta q$ and $\delta s$, and an arbitrary (infinitesimal) function $\epsilon (t)$.

The $s$-equation of motion, $\frac{\delta S}{\delta s}=\frac{\partial L}{\partial s}=0$ is of lower order in derivatives: the zeroth order for $s$ and first order for $q$. Expressing then $\dot q$ in terms of $p_q=\frac{\partial L}{\partial \dot q}$, we generically get the secondary constraint following from preservation of $p_s=0$ in the Hamiltonian approach\footnote{This is precisely because, for the canonical Hamiltonian taken as a function of coordinates, velocities and momenta, the surface of the definition of momenta is stationary with respect to velocities thus allowing us to get rid of them, modulo a possibility of several branches. Calculating the Poisson bracket $\{p_s, H\}$, note that, for $\delta s$ variation, the variation of $p_q {\dot q} - L$ at fixed $q$ and $p_q$ is equal to $\left(p_q - \frac{\partial L}{\partial {\dot q}}\right) \delta {\dot q} - \frac{\partial L}{\partial s} \delta s = - \frac{\partial L}{\partial s} \delta s$, the variation of $L$ at fixed $q$ and $\dot q$.}. Taking a gauge of $q=0$, we get an algebraic equation for $s$. Therefore, there is one gauge freedom and one fully constrained physical mode, and no dynamics at all. However, a gauge of $s=0$ leaves us with one half of a fake degree of freedom. Had we put $s=0$ directly into the action, the only equation would be $\frac{\delta S}{\delta q}=0$ which is second order in time derivatives of $q$. It is one degree of freedom; one half in the remnant gauge freedom, and another half in the new freedom for gauge-invariant variables.

This is fully in parallel with what we have seen above. Note also that everything is self-consistent, and the Lagrangian equations are, for sure, not fully independent. Indeed, the gauge invariance of the Lagrangian (\ref{meclag}) leads to an identity among the equations of motion:
$$(h-{\dot f}) \cdot \frac{\delta S}{\delta s} - f \cdot \frac{d}{dt} \frac{\delta S}{\delta s} +g \cdot \frac{\delta S}{\delta q} \equiv 0.$$
In other words, the $s$-equation is indeed a constraint from which the higher order $q$-equation follows. In gauge field theories, an analogous statement is also true, at least for the time development. Though, in absence of boundary conditions, the other equations might give more information on spatial shapes. For example, in electrodynamics $\partial_{\mu} \frac{\delta S}{\delta A_{\mu}} \equiv 0$.

{\bf A toy model}. In a simple model \cite{meHam} of $L=\frac12 ({\dot q} +s)^2$ the $s$-equation ${\dot q}+s=0$ is indeed stronger than the $q$-equation ${\ddot q}+\dot s=0$. There is no dynamical mode, one gauge freedom, so that $q$ can be taken as pure gauge, and one fully constrained physical mode represented by the condition of $s=-\dot q$. In the Hamiltonian language, $H_T=\frac12 p_q^2- s p_q +\lambda p_s,$ the primary constraint of $p_s=0$ entails the secondary one of $p_q=0$, and the Hamiltonian equations also reduce to ${\dot q}=-s$.

Putting a relaxation gauge of $s=0$ into the equations results in $\dot q =0$ which featues one half fake dynamical degree of freedom of the remnant gauge symmetry. Imposing it at a higher theoretical level does change the model. In Lagrangian formalism, it gives $L=\frac12 {\dot q}^2$ and therefore $\ddot q=0$. At the same time, $\utilde{\lambda} s$ added to the total Hamiltonian cancels any secondary constraint appearance and also leads to $\ddot q=0$. This is one degree of freedom, only one half of which is in the remnant gauge symmetry, while another half is a genuinely new content.

Note that a more general gauge of $s=f(q)$ also produces the same type of result. At the level of equations, it is an illusion of one half degree of freedom, ${\dot q} + f(q)=0$, due to the remnant gauge symmetry. However, when being done in a relaxation manner, it gets us ${\ddot q} + f^{\prime}(q) {\dot q}=0$. In other words, the stronger equation, the constraint, has been removed, and we end up with the weaker ${\ddot q} + {\dot s}=0$ equation only.

{\bf Adding spatial derivatives}. With spatial derivatives, the picture changes unless we wish to impose boundary conditions at spatial infinity. For illustration, let's assume a $(1+1)$-dimensional spacetime of Cartesian coordinates $(t,x)$ and a Lagrangian density ${\mathcal L}=\frac12 ({\dot q} +s^{\prime})^2$. The two equations, ${\ddot q}+{\dot s}^{\prime}=0$ and ${\dot q}^{\prime}+ s^{\prime\prime}=0$, are not independent but one is not strictly stronger than the other either. The only physical information is that the quantity ${\dot q} +s^{\prime}$ is constant. One might say, it is one half of a global degree of freedom.

The gauge of $s=0$, precisely as before, yields $\dot q=0$ at the level of equations, and $\ddot q=0$ when being imposed either in the Lagrangian density or as a hand-made constraint to the total Hamiltonian. The $x$-dependence is not restricted, and so it is an ultralocal type of dynamics. At the same time, the gauge of $q=0$ is no longer a full gauge fixing. In the same two versions of applying it, it leads to $s^{\prime}=0$ and $s^{\prime\prime}=0$ respectively, and with no restriction on time dependence, unless we go for boundary conditions and solve it as just $s=0$.

\section{The case of gravity}

 Finally, we would like to comment on relaxation of constraints in vacuum general relativity (GR). Before we do so, please note that one might say that general relativity is not a usual gauge theory, and therefore it is irrelevant for the gauge theory discussions. It is then possible to turn to other formulations such as MacDowell-Mansouri \cite{ggrMM} or new works on Lorentz gauge theory \cite{ggrn1,ggrn2,ggrn3}, or to mathematical considerations of teleparallel gravity \cite{ggrt1,ggrt2}. However, for our purposes, GR {\it is} a gauge theory, for it does enjoy a local symmetry when formally taken in terms of the metric tensor components as its variables.

In this case, there are immediately two different choices of variables for Hamiltonian description, the metric tensor components and ADM variables \cite{ADM}. One way is to use the metric tensor components {\`a} la Dirac \cite{gravDirac}, with the primary constraints of $\pi^{0\mu}=0$ after some transformation of the action functional. Relaxation gauge conditions of $g_{0\mu}=f_{\mu}(x)$ make the action, or the total Hamiltonian, produce then \cite{strgr} only the spatial part of Einstein equations $G^{ij}=0$. Therefore, it is indeed akin to adding an effective energy-momentum tensor with only non-spatial components $T^{0\mu}$ non-vanishing \cite{strgr}. Of course, it is subject to covariant conservation.

Roughly speaking, we have got an effective fluid without any pressure or shear stress. It actually reminds us of mimetic gravity \cite{mim1, mim2}. It would be interesting to see how far this analogy could go. We are not sure that one should be happy \cite{strcons} about such possible addition. In mimetic gravity, it can be seen as a problem. Indeed, a pressureless ideal fluid does generally develop caustic singularities. For a physical fluid, we just realise that the no-pressure, or an ideal-fluid approximation has failed. In mimetic gravity it becomes a fundamental feature though. Note in passing that we do not consider numerous modifications of mimetic gravity \cite{SVrev}.

Interestingly, in the paper \cite{strgr}, even though claiming the $T^{0\mu}$ addition, they discuss the Hamiltonian General Relativity in the ADM variables \cite{ADM}. In general, the simplest relaxation gauges, $N=f(x)$ and $N_i=f_i(x)$, produce then different relaxed theories because fixing the lapse and shift is not the same as fixing the non-spatial metric components. Variation of $g_{ij}$ with fixed $g_{0\mu}$ is not the same as variation of $g_{ij}$ with fixed $N$ and $N_i$. The point is that they imposed the synchronous gauge, and then the condition of $g_{00}=1$ and $g_{0i}=0$ is indeed equivalent to $N=1$ and $N_i=0$ since $g_{0i}=N_i$ and $g_{00}=N^2 - N_i N_j h^{ij}$ where $g_{ij}=-h_{ij}$. 

However, in general for fixing non-zero shifts, it is not the case. Indeed, in the purely metric component variables, the standard relaxation goes in terms of restricting the variations to $\delta g_{00}=\delta g_{0i}=0$, thus loosing the non-spatial part of Einstein equations. At the same time, the ADM-type relaxation would make the lapse and shift fixed, and that means $\delta g_{0i}=0$ and $\delta g_{00}=-N_i N_j \delta h^{ij}$ such that, in case of fixing a non-vanishing shift, we get a nontrivial combination of spatial equations with the temporal one, similar to what we have seen above when fixing a gauge of  $ A_0 + f(A_i)=0$ in the Maxwell action.

Even the Hamiltonian analysis itself gets some modification due to changing the variables to ADM. It makes some people claim that those variables are not good for a proper Hamiltonian analysis of general relativistic dynamics \cite{strange, morestrange}. In reality, there is no problem, and it is just a feature of any gauge theory where even the canonical Hamiltonian is not uniquely defined \cite{meHam}.

One can also make variations of the topic. For example, fixing just $N$ or $g_{00}$, without restricting any other component, are different things. Interestingly, if we do it as $g_{00}=1$ with no other constraint added by hand, the result is an effective energy-momentum tensor with $T^{00}$ only, reproducing their result for cosmology \cite{strin}. In the minisuperspace, there is no difference between $g_{00}$ and $N^2$ though.

If to go only for simple relaxation gauges \cite{strin}, i.e. fixing values of momentum-constrained variables without invoking non-trivial functions of other variables, then Dirac and ADM Hamiltonian pictures will mostly give different results. Of course, without this restriction, one can perfectly relate the two pictures to each other. However, it is also possible to consider gauge choices which are not that simple in either of these types of variables. For example, a very important option is $\det(g)=1$ or $N=\frac{1}{\sqrt{\det(h)}}$ which gives you the unimodular gravity as, for example, reviewed in the paper \cite{Kamen}.

\section{Discussion and Conclusions}

In every situation when the gauge hits twice, the gauge symmetry mixes the variable of lower time derivatives with other variables which otherwise behave normally. This is precisely why it hits twice. The primary constraint is not a generator of the Lagrangian gauge transformation by itself. Instead, its preservation in time requires yet another, secondary constraint, and only a specific combination of theirs then serves as a real gauge generator \cite{Castellani}.

If one immediately fixes a gauge condition of the relaxation type, i.e. puts a new constraint of second class with the primary constraint, then the system no longer needs any secondary constraint, for the time preservation of the given constraints is then ensured by specifying the values of Lagrange multipliers. In Lagrangian language, it is because we have removed the equation which used to be a constraint for the time evolution. Recently, the very same proposal, in the context of quantum cosmology, has appeared in a different but equivalent formulation \cite{Vasak}.

All in all, what happens in the relaxation procedure is that we choose to impose a gauge fixing condition before looking for the secondary constraint. Moreover, we insist on using an incomplete gauge fixing thus producing a remnant gauge freedom, and also some fully new dynamics if the condition was imposed before deriving the equations of motion. Being of the lower time-derivative order, the momentum-constrained variable has naturally got a time derivative of the gauge parameter in its gauge transformation, and therefore fixing the gauge in terms of such variable is always incomplete, whatever boundary condition we might have imposed at the spatial infinity. Hence, it is not really correct to treat the lapse $N$, or the vector potential $A_0$ component, as merely a gauge parameter \cite{Vasak, Sheikh}.

Gauge symmetry is an important concept in theoretical physics. Among other things, it allows us to write a nice action principle for equations of interaction fields. Therefore, removing it at the foundational level is a strange idea, to start with, let alone doing it with a gauge fixing which is necessarily incomplete even in perturbation theory. At the same time, an absolutely full gauge fixing would be difficult to naturally define in a field theory, and globally it is just impossible \cite{onGribov} for the physically interesting non-Abelian theories due to Gribov ambiguity\footnote{This is a topological obstruction \cite{onGribov}. Gribov himself found it \cite{theGribov} as a special case of remaining gauge freedom after imposing the Lorenz condition $\partial_{\mu}A_a^{\mu}=0$ in Euclidean time with boundary conditions at infinity.}.

In conclusion, we stress it again that the constraint relaxation idea \cite{strin, strgr, strelm, strcons} is a way of modifying the {\it classical} theory, and different versions of it can be found in research papers over many years \cite{Kamen}. However, we find it curious to see that it is precisely what happens when we decide to fix a gauge before finding secondary constraints and do it in a second-class way with the primary constraints. Therefore, it is similar to fixing a gauge in the approach with an extended Hamiltonian, done at another step of the Hamiltonian analysis though.

\end{document}